\documentclass[compact]{jetbrains_report}

\usepackage[style=authoryear,natbib=true,backend=biber]{biblatex}
\tcbuselibrary{breakable}
\tcbuselibrary{skins}

\usepackage{fancyvrb}
\usepackage{fvextra}
\usepackage{xspace}
\usepackage{comment}
\usepackage[capitalize,nameinlink]{cleveref}

\newcommand{\icaeagent}{\textsc{ICAE}\xspace}
\xspaceaddexceptions{-}

\makeatletter
\newcommand{\@acro@list}{}
\newcommand{\defacro}[2]{%
  \expandafter\newif\csname if@acro@#1@used\endcsname
  \expandafter\gdef\csname @acro@#1@short\endcsname{#1}%
  \expandafter\gdef\csname @acro@#1@long\endcsname{#2}%
  \g@addto@macro\@acro@list{\@acro@reset{#1}}%
}
\newcommand{\@acro@reset}[1]{%
  \global\csname @acro@#1@usedfalse\endcsname
}
\newcommand{\ac}[1]{%
  \@ifundefined{if@acro@#1@used}{%
    \PackageError{acronym}{Acronym '#1' not defined}{}%
  }{%
    \csname if@acro@#1@used\endcsname
      \csname @acro@#1@short\endcsname
    \else
      \csname @acro@#1@long\endcsname\ (\csname @acro@#1@short\endcsname)%
      \global\csname @acro@#1@usedtrue\endcsname
    \fi
  }%
}
\newcommand{\acp}[1]{%
  \csname if@acro@#1@used\endcsname
    \csname @acro@#1@short\endcsname s%
  \else
    \csname @acro@#1@long\endcsname s (\csname @acro@#1@short\endcsname s)%
    \global\csname @acro@#1@usedtrue\endname
  \fi
}
\newcommand{\acresetall}{\@acro@list}
\makeatother

\defacro{LLM}{Large Language Model}
\defacro{SE}{Software Engineering}
\defacro{ICAE}{In-Context Autoencoder}
\defacro{LoRA}{Low-Rank Adaptation}
\defacro{RoPE}{Rotary Positional Embeddings}
\defacro{SFT}{Supervised Fine-Tuning}
\defacro{SQuAD}{Stanford Question Answering Dataset}
\defacro{BLEU}{Bilingual Evaluation Understudy}
\defacro{DMCA}{Digital Millennium Copyright Act}
\defacro{URL}{Uniform Resource Locator}
\defacro{QA}{Question Answering}
\defacro{PR}{Pull Request}

\title{On Problems of Implicit Context Compression\\ for Software Engineering Agents}

\jbmetadata[s]{Date}{April 2026}

\jbauthor{Kirill Gelvan}{1,2}
\jbauthor{Igor Slinko}{1}
\jbauthor{Felix Steinbauer}{2}
\jbauthor{Egor Bogomolov}{1}
\jbauthor{Florian Kofler}{2}
\jbauthor{Yaroslav Zharov}{1}
\jbaffiliation{1}{JetBrains Research}
\jbaffiliation{2}{Technical University of Munich, Germany}
\jbcorresponding{kirill.gelvan@jetbrains.com}

\begin{document}

\maketitle

\begin{jbabstract}
LLM-based Software Engineering agents face a critical bottleneck: context length limitations cause failures on complex, long-horizon tasks.
One promising solution is to encode context as continuous embeddings rather than discrete tokens, enabling denser information storage.
We apply the recently proposed In-Context Autoencoder for this purpose.
While the method performs well on single-shot common-knowledge and code-understanding tasks, our experiments demonstrate that it fails on multi-step agentic coding tasks.
In this paper, we explore this phenomenon and discuss possible factors contributing to this failure.
\end{jbabstract}
\acresetall

\section{Introduction} \label{sec:introduction}

\ac{SE} is increasingly automated by \ac{LLM}-based agents that write and fix code~\citep{yang2024swe,jimenez2023swe}.
These agents interact with a development environment over multiple turns to perform complex tasks, such as feature implementation or bug fixing.
At each step, the agent takes observations from the environment (e.g., output of a previous command, or input from the user), prompts an \ac{LLM} to reason about the next action given the history of previous actions and observations, and then executes this action.
Different \ac{LLM}-based agents vary in architectures, but the described framework, called ReAct~\citep{yao2022react}, is one of the bases on which many are built.

However, the effectiveness of such agents is limited by the amount of information they can process~\citep{shi2025look}.
This limitation comes from the fixed context window of the underlying models.
As agents accumulate observations from tools (e.g., file contents, error logs, and command outputs), they quickly exhaust this window~\citep{kang2025acon,labate2025solving}.
While modern hardware and optimizations allow processing millions of tokens, models fail to function beyond their training context length.

This hard limit leads to a halt in operation or a steep drop in quality when ignored.
Additionally, as the context grows, models suffer from the ``needle in a haystack'' phenomenon, where the accuracy degrades as relevant information becomes buried in long sequences~\citep{liu2024lost}.
As \cite{badertdinov2025swe} show, a single SWE-bench-like issue requires consuming over a million tokens on average.

To address this bottleneck, various solutions have been proposed.
Some efforts focus on training models with massive context windows~\citep{comanici2025gemini} or modifying architectures with mechanisms like Infini-attention~\citep{munkhdalai2024leave}.
Others employ mechanisms to mitigate \ac{RoPE}~\citep{su2021roformer} extrapolation issues, such as NoPE~\citep{kazemnejad2023impact} or DroPE~\citep{gelberg2025extending}, or strategies that explicitly drop or compress earlier parts of the interaction trajectory to save space~\citep{pan2024llmlingua,wang2024natural}.

Implicit context compression represents a particular compression approach where text is encoded into continuous embeddings rather than discrete tokens~\citep{dai2025pretraining}.
This approach operates under the assumption that dense real-valued vector representations can hold more information than discrete tokens~\citep{morris2023text}.
Consequently, encoders are trained to reduce the number of required ``attention slots'' by generating fewer embeddings than there were tokens in the input.
Recent works demonstrate that reasoning chains can be effectively compressed into a few continuous embeddings~\citep{hao2024coconut,shen2025codi,xu2025softcot}, suggesting the potential for long agentic workflows.

\textbf{In this work,} we investigate implicit context compression for SWE agents using the \ac{ICAE} approach~\citep{ge2023context} (detailed in~\Cref{sec:method}).
This method proposes an encoder-decoder architecture to compress the input context into continuous embeddings.
We adapt \ac{ICAE} for \ac{SE} agent tasks by fine-tuning on multi-step trajectories from the SWE-Smith dataset~\citep{yang2025swe} and evaluate its performance on the SWE-bench Verified benchmark~\citep{chowdhury2024swebenchverified}.
We observe a surprising lack of generalization from single-step tasks to multi-step trajectories, investigate possible reasons with additional experiments in \Cref{sec:experiments}, and propose two potential explanations in \Cref{sec:discussion}.

\section{Method} \label{sec:method}

This section details the adaptation of implicit context compression to agentic \ac{SE} tasks, based on the In-Context Autoencoder~\citep{ge2023context}.
We outline the original \ac{ICAE} architecture, its standard pretraining and fine-tuning procedures, and our modifications to them.
We then present our modifications for the agentic domain.

\textbf{Architecture.} The \ac{ICAE} architecture consists of two decoder-only transformers: a trainable encoder and a frozen decoder, both initialized from the same pretrained foundation model.
The encoder is trained with \ac{LoRA}~\citep{hu2022lora} to compress long contexts into a fixed sequence of continuous embeddings, called \textit{memory tokens}.
The decoder attends to these memory tokens instead of the original text to generate output.
Unlike other approaches~\citep{bulatov2022recurrent, jaegle2021perceiver}, the decoder remains frozen, preventing catastrophic forgetting~\citep{mccloskey1989catastrophic}.
We replace the Llama-2-7B base model~\citep{touvron2023llama} with Qwen3-8B~\citep{yang2025qwen3}, as the latter is reported to perform better on agentic tasks.

\textbf{Pretraining.}
The goal of pretraining is to initialize the compression capability using a massive text corpus. \citet{ge2023context} originally used The Pile~\citep{gao2020pile}.
Training employs a 50/50 mix of autoencoding (reconstructing the input text verbatim) and language modeling (continuation of the compressed text) objectives.
To make the decoder aware of the task, the compressed memory tokens are appended with a special task-signaling token.
For both objectives, the decoder generates the output based on the compressed context, but \textit{only the encoder weights} are optimized.
This forces the encoder to learn representations that preserve semantic information in a format the frozen decoder can understand.
Due to the DMCA takedown\footnote{\url{https://en.wikipedia.org/wiki/The_Pile_(dataset)\#Training_on_copyrighted_works_or_derivatives}} of The Pile dataset, we instead use SlimPajama-6B~\citep{weber2024redpajama} for pretraining.

\textbf{Fine-tuning.}
The next stage fine-tunes the encoder for specific downstream tasks (originally for Question Answering).
The model receives a long context, which the encoder compresses into memory tokens.
The user's question is kept in its original discrete token form and appended to the memory tokens.
The frozen decoder then attends to both the memory tokens representing the document and the discrete tokens of the question to generate the answer.
Gradients from the answer generation are backpropagated to update the encoder, optimizing the compressed representation for the task.

\textbf{Agentic Fine-Tuning.}
We adapt \ac{ICAE} for agentic workflows by modifying the task structure.
For fine-tuning on agentic trajectories, we compress only observations longer than 256 tokens and compute backpropagation only for the latest compressed observation.
Due to memory constraints, all previous observations are kept cached as memory tokens until the trajectory is finished.
Short observations, actions, and the system prompt are preserved in their original discrete token form.
The model is trained to predict the next action (tool call) based on a history containing compressed observations.
Following~\citet{zhao2024position}, we employ position ID manipulation to minimize the effective distance between memory tokens and the current prompt, facilitating better attention mechanisms in \ac{RoPE}-based models.
\Cref{app:icae-examples} further illustrates the training and inference process.

\section{Experiments} \label{sec:experiments}

We evaluate our modified version of \ac{ICAE} across three distinct scenarios for context compression. The first one tests the model's ability to retrieve explicit facts from compressed natural language using \textbf{SQuAD}~\citep{rajpurkar2016squad}.
The second assesses whether the compression preserves the high-fidelity details required for code comprehension using \textbf{RepoQA}~\citep{liu2024repoqa}.
The third evaluates the model's capacity to plan and execute multi-step \ac{SE} tasks using \textbf{SWE-bench Verified}~\citep{chowdhury2024swebenchverified}.

To assess performance across these scenarios, we compare three configurations:
(i) \textbf{Base} -- Qwen3-8B~\citep{yang2025qwen3} without fine-tuning or compression,
(ii) \textbf{SFT} -- Qwen3-8B fine-tuned on the respective task without compression, and
(iii) \textbf{\icaeagent} -- our modification, with the encoder pretrained and then fine-tuned as described in~\Cref{sec:method}.
We disable thinking for all experiments to ensure comparable results.
All quantitative results are presented in \Cref{tab:main} and discussed further in this Section.
For the sake of space, we share all training hyper-parameters in~\Cref{app:train-hyp} and detailed experiment statistics in~\Cref{app:detailed-runs}.

\newcommand{\pdiff}[1]{\,\raisebox{1pt}{\textcolor{green!65!black}{\scriptsize(+#1)}}}
\newcommand{\ndiff}[1]{\,\raisebox{1pt}{\textcolor{red}{\scriptsize\textbf{(-#1)}}}}
\begin{table}[h]
  \centering
  \small
  \setlength{\tabcolsep}{4pt}
  \begin{tabular*}{\textwidth}{@{\extracolsep{\fill}}lcllllll@{}}
  \toprule
  \multirow{2}{*}{\textbf{Model}} & \multirow{2}{*}{\textbf{\shortstack{Compress.\\Rate}}} & \multicolumn{2}{c}{\textbf{SQuAD}} & \multicolumn{2}{c}{\textbf{RepoQA}} & \multicolumn{2}{c}{\textbf{SWE-bench Verified}} \\
  \cmidrule(lr){3-4} \cmidrule(lr){5-6} \cmidrule(lr){7-8}
  & & \multicolumn{1}{l}{\textbf{BLEU}} & \multicolumn{1}{l}{\textbf{EM}} & \multicolumn{1}{l}{\textbf{BLEU}} & \multicolumn{1}{l}{\textbf{Pass@0.8}} & \multicolumn{1}{l}{\textbf{BLEU}$_{\text{ref}}$} & \multicolumn{1}{l}{\textbf{Resolved Issues}} \\
  \midrule
  Base & $1\times$ & 0.67 & 0.54 & 0.81 & 0.65 & 0.48 & \underline{19} \\[0.25em]
  SFT & $1\times$ & \textbf{0.75}\pdiff{0.08} & \textbf{0.70}\pdiff{0.16} & \textbf{0.90}\pdiff{0.09} & \textbf{0.85}\pdiff{0.20} & \textbf{0.55}\pdiff{0.07} & \textbf{86}\pdiff{67} \\
  \midrule
  \icaeagent & $4\times$ & \underline{0.73}\pdiff{0.06} & \underline{0.67}\pdiff{0.13} & \underline{0.87}\pdiff{0.06} & \underline{0.69}\pdiff{0.04} & \underline{0.51}\pdiff{0.03} & 7\ndiff{12} \\
  \bottomrule
  \end{tabular*}
  \caption{Performance report.
  Best results in \textbf{bold}, second best \underline{underlined}.
  Performance differences with Base model are reported in parentheses.
  SFT uses LoRA for SQuAD and RepoQA, but full fine-tuning for SWE-bench Verified.}
  \label{tab:main}
\end{table}

\textbf{SQuAD.}
In SQuAD, each sample consists of three parts: a question, a context paragraph, and an answer.
Given the question and the context paragraph, the model is expected to provide the correct answer.
We fine-tune the encoder on the training set to compress the context paragraph.
The fine-tuning target is for the decoder to take as input the uncompressed question and compressed context and predict the answer.
We evaluate performance on the standard validation set, reporting \ac{BLEU} scores~\citep{papineni2002bleu} and the Exact Match (EM).
We observe that \icaeagent significantly improves over the baseline ($p < 0.001$), while underperforming the SFT checkpoint.
This confirms that compression aids in extracting knowledge from general text but has a minor penalty compared to full-context access.

\textbf{RepoQA.}
We utilize the ``Searching Needle Function'' task from the RepoQA dataset to test code retrieval.
In this task, the model is required to retrieve a specific function from the whole code repository given in the context, based on a short natural language description.
The fine-tuning target is to take as input the uncompressed instructions and function description and compressed code to predict the target function.
We use a context length of 8192 tokens to verify high-fidelity retrieval. We follow the authors and use \ac{BLEU} and Pass@0.8 metrics averaged across all five programming languages in the benchmark.
Pass@0.8 measures the percentage of generations that achieve a \ac{BLEU} score of at least 0.8 with the ground truth, and is calculated after extracting the code block from the generated string and removing comments.
Across 5 runs, we find no statistically significant difference between \icaeagent and Base, confirming that compression maintains retrieval capabilities, whereas SFT yields substantially larger gains on the downstream metric.

\textbf{SWE-bench Verified.}
We use the whole dataset of 500 \ac{SE} tasks as the evaluation set.
This dataset consists of 12 repositories, issue texts for each repository, and a set of tests for each issue.
A successful task resolution is registered if all corresponding tests pass after the proposed code patch is applied.
The fine-tuning target is for the decoder to predict the next action given the series of uncompressed actions and compressed observations.
We use expert successfully resolved agentic trajectories from the SWE-smith dataset~\citep{yang2025swe} for fine-tuning, reserving 300 trajectories as a test set.
Due to the complexity of the task, the SFT uses full fine-tuning instead of \ac{LoRA}.
However, to remain consistent with the original paper, our \icaeagent continues to use the same training hyperparameters as before.
Since this benchmark doesn't provide ground truth actions, as an intermediate metric of quality, we measure \ac{BLEU}$_{\text{ref}}$, which stands for \ac{BLEU} with respect to the expert trajectories from the test subset.
As the downstream metric, we use the number of resolved issues.
The \ac{BLEU}$_{\text{ref}}$ metric follows the usual trend --- \icaeagent outperforms the base model and is outperformed by SFT.
However, the downstream metric results are drastically different --- the \icaeagent significantly underperforms both the Base model (12 fewer issues solved, $p=0.013$), and the SFT checkpoint ($79$ fewer issues solved).

While the method fails to increase the number of resolved issues, it successfully extends the effective context window size.
The model was trained with a $4\times$ compression rate, but the effective compression rate, which depends on the dataset characteristics and which parts are selected for compression, is lower. The effective compression is $1.46\times$ on SQuAD, $3.74\times$ on RepoQA, and on $2.0\times$ SWE-bench Verified.
We note that the effective compression rate does not correlate with the downstream performance.
Practically, on SWE-bench Verified, the method allows for $40\%$ longer trajectories ($113$ vs.\ $81$ steps on average; see \Cref{app:trajectory-length}) and $10\%$ faster generation time (incl. compression).

\section{Discussion} \label{sec:discussion}

The results show that compression provides a similar or slightly increased downstream performance for single-shot experiments.
However, in the multi-step setting, the \icaeagent underperforms the base model.
Since the decoder wasn't trained, we can attribute the problem to the compression mechanism.
We identify two hypotheses to explain the failure of implicit compression in the agentic setting: error accumulation and failure to account for long-range dependencies.

\textbf{Reconstruction Fidelity and Error Accumulation} concerns the fidelity of the compressed representation.
Through qualitative analysis, we observed examples where URLs or file paths were hallucinated during reconstruction, such as replacing \texttt{swe-agent.com/latest/} with \texttt{swe-agent.com\textit{/agent/}latest/}, invalidating the output.
In single-step tasks, such errors are isolated to a single output.
However, in multi-step agentic workflows, these errors compound.

\textbf{Information Preservation for Future Steps} relates to the limitations of the compression objective during fine-tuning.
In our fine-tuning setup, at timestep $k$, all the observations $1,\dots,k-1$ are compressed.
However, only the last encoder instance (the weights that compressed the observation $k-1$) is optimized.
Hence, the training signal flows through only this single step, and the model has no incentive to preserve the information needed for subsequent steps.
Nevertheless, since each step involves a full pass through two \acp{LLM}s (encoder and decoder), addressing this would make the training of such architectures computationally very challenging for long sequences.

To confirm or refute these hypotheses, we propose future work to experiment with a more granular selection of steps for compression.
In this work, we selected observations to compress based on their length; however, further investigation of failure patterns may help to pinpoint the problem.
For example, compressing only the last $k$ steps may test the \textbf{information preservation} hypothesis, and randomly compressing $k\%$ of steps may estimate the compression losses regarding \textbf{reconstruction fidelity}.

\section{Conclusion} \label{sec:conclusion}

In this paper we evaluated implicit context compression for \ac{LLM}-based SWE agents.
While \icaeagent successfully improves efficiency (40\% longer trajectories, 10\% faster generation) and outperforms the baseline on single-shot tasks (general and coding), it fails in the multi-step agentic \ac{SE} setup.
Specifically, we observe a significant performance degradation on SWE-bench Verified compared to the baseline model.

Our analysis suggests two possible contributing factors: insufficient reconstruction fidelity, which compounds errors over time, and a failure to selectively preserve crucial information for future steps due to single-step training objectives.
These challenges are non-trivial to resolve and require a more involved modeling and training approach than anticipated (e.g., full-trajectory reinforcement learning).
We suggest that future work should focus on identifying the specific challenges in adapting implicit compression to agentic tasks.

\section*{LLM Usage Statement}
We utilized LLMs for initial drafting and grammar polishing.
However, the substantial writing, all experimental design and analysis remain the original contribution of the authors.

\section*{Reproducibility Statement}
\begin{sloppypar}
To support reproducibility, we release our code and model checkpoints.
The source code for the \ac{ICAE} implementation, data processing, and evaluation scripts is available at \url{https://github.com/AnonForConference/paper1-code-0126}.
The model checkpoints, including the pretrained encoder and the agentic fine-tuned model, are hosted at \url{https://huggingface.co/AnonForConference/paper1-model-0126}.
\end{sloppypar}

\newpage
\printbibliography

\newpage
\appendix
\crefalias{section}{appendix}
\crefalias{subsection}{appendix}

\section{Training Hyperparameters}
\label{app:train-hyp}

\Cref{tab:hyperparams} summarizes the training hyperparameters.

\begin{table}[h]
  \centering
  \small
  \begin{tabular}{lcc}
  \toprule
  \textbf{Parameter} & \textbf{Pretraining} & \textbf{Fine-tuning} \\
  \midrule
  Base Model & Qwen3-8B & Qwen3-8B \\
  Optimizer & AdamW & AdamW \\
  Learning Rate & $1 \times 10^{-4}$ & $5 \times 10^{-5}$ \\
  Batch Size & 1 & 1 \\
  Gradient Accumulation & 8 & 1 \\
  \ac{LoRA} Rank & 128 & 128 \\
  \ac{LoRA} $\alpha$ & 32 & 32 \\
  Target Modules & q\_proj, v\_proj & q\_proj, v\_proj \\
  Warmup Steps & 300 & 300 \\
  \bottomrule
  \end{tabular}
  \caption{Training hyperparameters for pretraining and fine-tuning stages. We utilize the same hyperparameters across all fine-tuning datasets, varying only the number of training steps: 100,000 for Pretraining, 10,000 for SQuAD, 4,000 for RepoQA, and 150,000 for SWE-bench Verified.}
  \label{tab:hyperparams}
\end{table}

Here we present the agent's scaffolding details. The agent interacts with the environment using one of the three tools:
\begin{enumerate}
\item \texttt{bash}: standard shell interface for running commands;
\item \texttt{submit}: submits the final patch for evaluation; and
\item \texttt{str\_replace\_editor}: stateful file editor supporting view, create, str\_replace, insert, and undo\_edit operations.
\end{enumerate}

The \texttt{str\_replace\_editor} requires exact line matching for replacements, ensuring deterministic edits.
This precise control is essential for making targeted code changes but also means that small reconstruction errors in compressed observations can lead to failed edit attempts.

This interaction protocol is adopted from SWE-smith, where the complete system prompt and tool specifications can be found.

\section{Detailed Multi-Run Experiment Results}
\label{app:detailed-runs}

\begin{figure}[ht]
    \centering
    \includegraphics[width=0.95\textwidth]{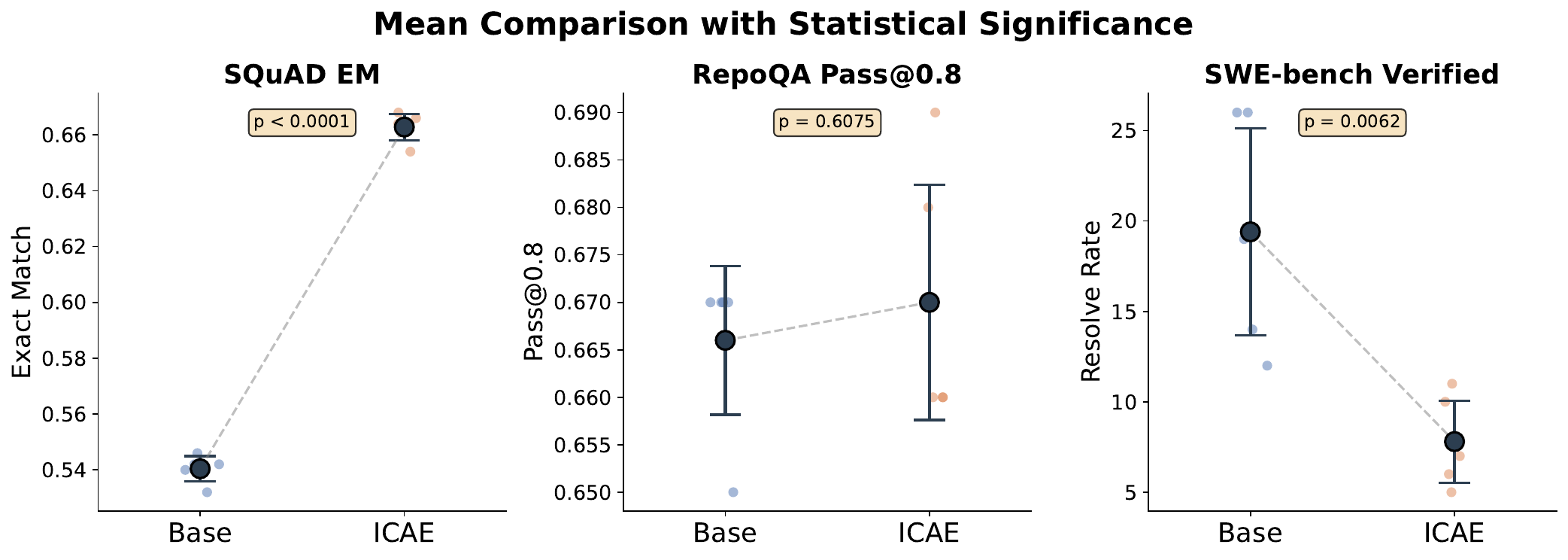}
    \caption{Comparison of Base and \icaeagent performance across three benchmarks.
 Each subplot shows mean values with 95\% confidence intervals (error bars) and
 individual run results (scattered points). Statistical significance was assessed using
  independent samples t-tests. \icaeagent significantly outperforms Base on SQuAD EM
 ($p < 0.0001$), shows no significant difference on RepoQA Pass@0.8 ($p = 0.6075$), and
 performs significantly lower on SWE-bench Verified resolve rate ($p = 0.0062$).}
   \label{fig:statistical-comparison}
\end{figure}

In this section, we list more detailed results of the multiple runs for our experiments and describe the statistics.
To confirm statistical significance, we performed a two-sample Welch t-test across 5 independent runs.
The results, including the statistical significance are shown in~\Cref{fig:statistical-comparison}

\newpage
\section{Trajectory Length Comparison}
\label{app:trajectory-length}

\begin{figure}[ht]
    \centering
        \centering
        \includegraphics[width=0.7\textwidth]{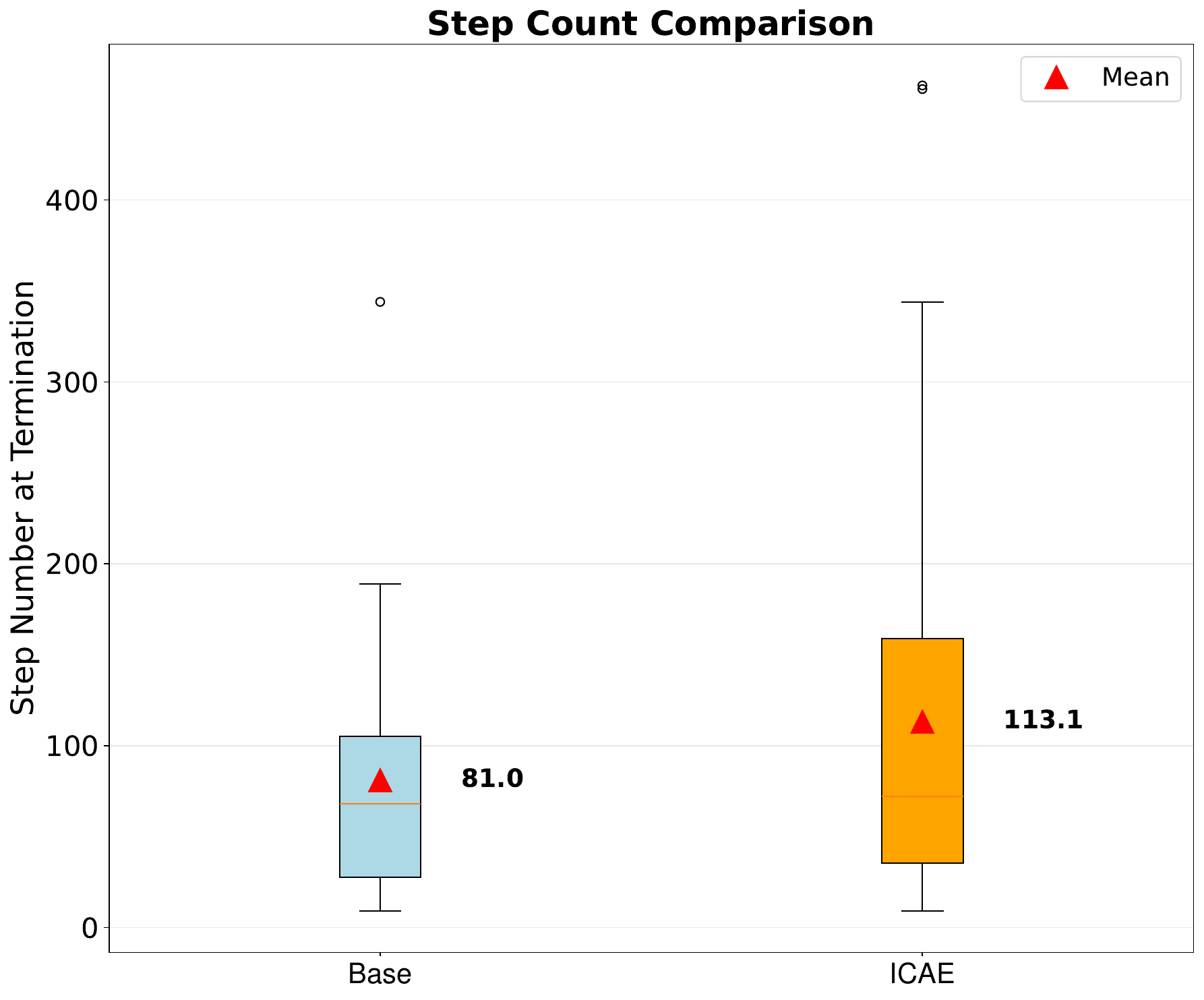}
    \caption{Comparison of trajectory lengths of \icaeagent and Base model. Comparison is performed at termination (context limit 32k tokens, no step limit).
    Triangles indicate mean values across models.}
    \label{fig:boxplot-stepcount}
\end{figure}

\Cref{fig:boxplot-stepcount} illustrates the trajectory length distribution.
The compressed agent (\icaeagent) executes significantly more steps before reaching the context limit compared to the uncompressed baseline.

\section{\icaeagent Training and Inference Details}
\label{app:icae-examples}

In this section, we further detail the \icaeagent training and inference.
The following example demonstrates a hand-crafted scenario, closely related to SWE-bench. This scenario was designed to test the model's ability to comprehend and utilize decompressed data from its memory. The environment for this sample was manually simulated by one of the authors. In this particular example, the model is able to run the secret command that has only been observed in the compressed form of memory tokens.

\begin{figure}[hbt]
\centering
    \begin{subfigure}[b]{\textwidth}
        \centering
        \includegraphics[width=\textwidth]{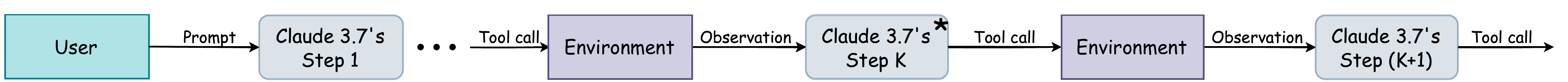}
        \caption{Overview of the agentic trajectory generation process used for fine-tuning data. The step marked with asterisk (Step $k$) is selected for compression because the previous observation has length of more than 256 tokens.}
        \label{fig:icae-agent-training-step-selection}
    \end{subfigure}
    \begin{subfigure}[b]{\textwidth}
        \centering
        \includegraphics[width=\textwidth]{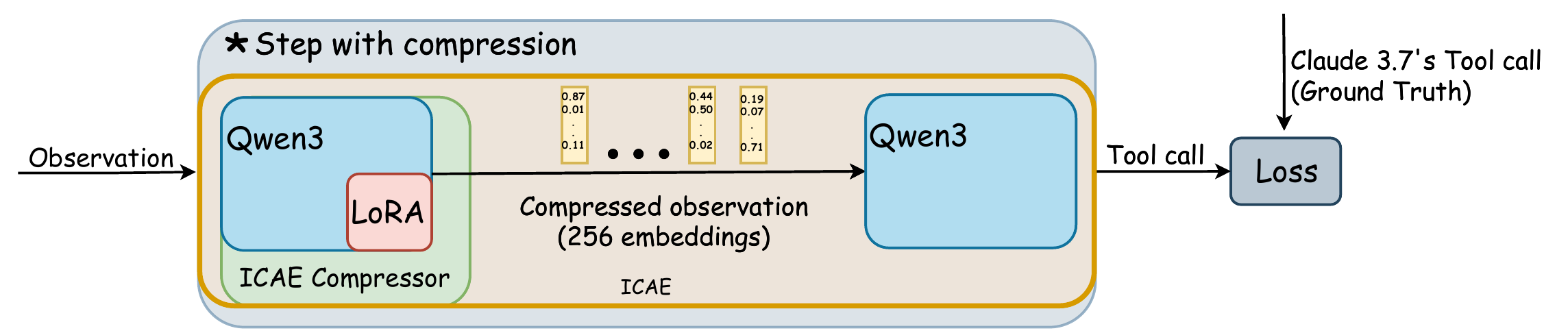}
        \caption{A single fine-tuning step for the agentic \ac{ICAE}. Observation from the previous step ($k-1$) is passed through ICAE-encoder, and then the next tool call is generated by ICAE-decoder. The loss is calculated between the ground truth and predicted tool call. The optimization is then applied only to the LoRA weights in the current ICAE-encoder.}
        \label{fig:icae-agent-training-step}
    \end{subfigure}
    \caption{Training of ICAE-encoder for the agentic setup. The step selection is illustrated in subfigure (a), and the optimization step is detailed in subfigure (b).}
     \label{fig:icae-agent-training}
\end{figure}

\begin{figure}[ht]
    \centering
    \begin{tcolorbox}[
        enhanced, sharp corners,
        boxrule=0.6pt, colframe=jb-gray!25, colback=jb-code-bg,
        left=8pt, right=8pt, top=6pt, bottom=6pt,
    ]
    \begin{Verbatim}[breaklines=true,obeytabs=false,breaksymbol={},breakindent=0pt,
                     fontsize=\small,commandchars=\\\{\}]
\textbf{\textcolor{jb-main}{User}}    <uploaded_files>/mnt/shared-fs/swe-agent-distillation</uploaded_files>
          I've uploaded a python code repo. Consider the \ac{PR} description:
          \colorbox{yellow!25}{Your task is to read README.md and run the commands for the setup.}
          \colorbox{yellow!25}{when i run them i get errors.}
          <fixed_part_of_the_first_user_message>*
\textbf{\textcolor{jb-teal}{Agent}}   $ str_replace_editor --command view --path /mnt/shared-fs/.../README.md
\textbf{\textcolor{jb-main}{User}}    \colorbox{yellow!25}{<MEMORY_TOKEN> x 512}
\textbf{\textcolor{jb-teal}{Agent}}   $ cat /mnt/shared-fs/swe-agent-distillation/requirements.txt
\textbf{\textcolor{jb-main}{User}}    There is no file named requirements.txt
\textbf{\textcolor{jb-teal}{Agent}}   $ str_replace_editor --command view --path /mnt/shared-fs/swe-agent-distillation
\textbf{\textcolor{jb-main}{User}}    You have no access to this directory.
\textbf{\textcolor{jb-teal}{Agent}}   $ ls -la /mnt/shared-fs/swe-agent-distillation
\textbf{\textcolor{jb-main}{User}}    You have no access to this directory. Call a command from README.md to install.
\textbf{\textcolor{jb-teal}{Agent}}   $ str_replace_editor --command view --path /mnt/shared-fs/.../README.md
\textbf{\textcolor{jb-main}{User}}    \colorbox{yellow!25}{<MEMORY_TOKEN> x 512}
\textbf{\textcolor{jb-teal}{Agent}}   \colorbox{yellow!25}{$ pip install -r /mnt/shared-fs/swe-agent-distillation/.my_secret_requirements.txt}
    \end{Verbatim}
    \end{tcolorbox}
    \caption{*\texttt{<fixed\_part\_of\_the\_first\_user\_message>} follows the SWE-smith system prompt format and is omitted for brevity.}
    \label{fig:icae-poc}
\end{figure}

\end{document}